%% file: main.tex
\documentclass[sigconf]{acmart}

\acmConference[arXiv Preprint]{}{2025}{}
\acmBooktitle{}
\renewcommand\footnotetextcopyrightpermission[1]{} 

\usepackage{algorithmic}
\usepackage{textcomp}
\usepackage{multirow}
\usepackage{stfloats}
\usepackage{subcaption}
\usepackage{csquotes}
\usepackage{soul}
\usepackage{listings}
\usepackage{framed}          

\definecolor{codegreen}{rgb}{0,0.6,0}
\definecolor{codegray}{rgb}{0.5,0.5,0.5}
\definecolor{codepurple}{rgb}{0.58,0,0.82}
\definecolor{backcolour}{rgb}{1,1,1}
\definecolor{MyBlue}{rgb}{0.0,0.5,1.0}
\definecolor{MyRed}{rgb}{1.0,0.0,0.0}

\colorlet{acmboxrule}{gray!75!black}
\colorlet{acmboxbg}{gray!5}
\makeatletter
\newenvironment{acmbox}{%
  \MakeFramed{\advance\hsize-\width \FrameRestore}%
}{\endMakeFramed}

\definecolor{acmTitleBG}{HTML}{1F4B99}
\definecolor{acmTitleFG}{HTML}{FFFFFF}
\definecolor{acmRule}{HTML}{1A2A4A}
\definecolor{acmBG}{HTML}{F4F7FF}

\newenvironment{acmTitledBox}[1]{%
  \MakeFramed{\advance\hsize-\width \FrameRestore}%
  \noindent
  \colorbox{acmTitleBG}{%
    \parbox{\dimexpr\linewidth-2\fboxsep\relax}{\bfseries\color{acmTitleFG}#1}%
  }%
  \par\smallskip
}{\endMakeFramed}
\makeatother

\begin{document}

\title{Assertion Messages with Large Language Models (LLMs) for Code}







\author{Ahmed Aljohani}
\affiliation{%
  \institution{University of North Texas}
  \city{}
  \state{}
  \country{}}
\email{ahmedaljohani@unt.edu}

\author{Anamul Haque Mollah}
\affiliation{%
  \institution{University of North Texas}
  \city{}
  \state{}
  \country{}}
\email{anamulhaque.mollah@unt.edu}

\author{Hyunsook Do}
\affiliation{%
  \institution{University of North Texas}
  \city{}
  \state{}
  \country{}}
\email{hyunsook.do@unt.edu}

\begin{abstract}
Assertion messages significantly enhance unit tests by clearly explaining the reasons behind test failures, yet they are frequently omitted by developers and automated test-generation tools. Despite recent advancements, Large Language Models (LLMs) have not been systematically evaluated for their ability to generate informative assertion messages. In this paper, we introduce an evaluation of four state-of-the-art Fill-in-the-Middle (FIM) LLMs— Qwen2.5-Coder-32B, Codestral-22B, CodeLlama-13b-hf, and StarCoder—on a dataset of 216 Java test methods containing developer-written assertion messages. We find that Codestral-22B achieves the highest quality score of 2.76 out of 5 using a human-like evaluation approach, compared to 3.24 for manually written messages. Our ablation study shows that including descriptive test comments further improves Codestral's performance to 2.97, highlighting the critical role of context in generating clear assertion messages. Structural analysis demonstrates that all models frequently replicate developers' preferred linguistic patterns. We discuss the limitations of the selected models and conventional text evaluation metrics in capturing diverse assertion message structures. Our benchmark, evaluation results, and discussions provide an essential foundation for advancing automated, context-aware generation of assertion messages in test code. A replication package is available at \textcolor{blue}{\url{https://doi.org/10.5281/zenodo.15293133}}.
\end{abstract}


\maketitle

\input{Sections/Introduction}

\input{Sections/RelatedWork}
\input{Sections/ResearchMethod}

\input{Sections/StudyResults}

\input{Sections/Study_Implication}

\input{Sections/ThreatsToValidity}

\input{Sections/Conclusion}

\bibliographystyle{ACM-Reference-Format}
\bibliography{Bib/bib}

\end{document}

%% file: Sections/Introduction.tex
\section{Introduction}
\label{sec:Introduction}

Unit testing-- is a crucial practice in software development that ensures the functionality correctness of individual components of a software system. At the heart of unit tests are assertion statements (e.g, Listing~\ref{lst:example} line 4), which verify whether the actual output of a program matches the expected outcome \cite{khorikov2020unit}. A well-constructed assertion not only determines the success or failure of a test but also provides valuable feedback to developers through assertion messages—descriptive texts that explain why an assertion fails \cite{peruma2024rationale}. These messages (e.g, Listing~\ref{lst:example} - \texttt{"Password dose not match"}) play a significant role in improving test code understandability, readability, and overall quality.

\begin{lstlisting}[basicstyle=\small, caption= Test Code Example, label=lst:example]
public void testCheckPassword() {
   String e = "Abc_123!";
   String a =checkPassword("Abc_123!");
   assertEquals("Password dose not match", e, a);}
\end{lstlisting}

Despite their importance, assertion messages are often overlooked in test code. Studies have shown that developers rarely include custom messages in assertion methods~\cite{takebayashi2023exploratory}. Approximately $ \approx 6\%$ of 31K test methods contain at least one assertion. This limitation is also common among automated test-generation techniques, including search-based approaches like EvoSuite~\cite{fraser2011evosuite} and transformer-based methods like AthenaTest~\cite{tufano2020unit}. For example, EvoSuite often generates multiple assertions per test method without descriptive messages, making the tests challenging to understand~\cite{panichella2022test}. Similarly, AthenaTest, despite being trained on developer-written tests, frequently generates assertions lacking informative messages~\cite{aljohani2024fine}. Moreover, existing automated tools for generating assertion statements primarily focus on verifying correctness or producing expected outcomes, but they rarely emphasize generating informative assertion messages~\cite{zamprogno2022dynamic, watson2020learning, he2024empirical, primbs2025assert5, zhang2025exploring} (See Section~\ref{sec:RelatedWork} for details). This lack of emphasis can lead to what the software engineering community refers to as a test smell\cite{bavota2015test}—specifically, \textit{Assertion Roulette}~\cite{aljedaani2021test}—where multiple assertions without clear failure messages complicate debugging efforts and increase developers' manual inspection workload. 

Recently, large language models (LLMs) have been explored for test code enhancement~\cite{wang2024software}. However, even these advanced models often fail to generate descriptive assertion messages~\cite{siddiq2024using, ouedraogo2024test} \textit{unless} explicitly guided via detailed prompt engineering techniques~\cite{deljouyi2024leveraging}.
While prior research, such as Deljouyi et al.~\cite{deljouyi2024leveraging}, leveraged LLMs to enhance test code through assertion generation messages, they did not explicitly evaluate how effectively LLMs can produce assertion messages. Peruma et al.~\cite{peruma2024rationale} highlighted developers' preference for assertion messages that (i) clearly describe test failures (expected versus actual outcomes), (ii) are textual and human-readable, and (iii) accurately reflect the failure's cause.

Given the ability of code LLMs to understand and generate human-readable text in tasks such as code summarization\cite{lu2021codexglue, sun2024source}, it is worth investigating whether they can produce assertion messages as effective as those written by developers.

We formed our motivation into research questions as follows:
\begin{enumerate} 
\item \textbf{RQ1 (Performance):} How do LLM-generated assertion messages compare to manually written ones?\\
We evaluate LLMs against developer-written messages and conduct an ablation study to assess the impact of additional contextual information (e.g., descriptive test comments) on generation quality.
\item \textbf{RQ2 (Structure and Linguistics):} How do the structural and linguistic patterns  of LLM-generated assertion messages align with developers-written assertion?\\
We examine the composition and phrasing of generated messages to determine their consistency with developer conventions and preferences.
\end{enumerate}

To investigate these research questions, we utilize a dataset of Java test methods extracted from 20 Java projects, filtering only those with manually written assertion messages as ground truth. We then prompt LLMs to generate assertion messages using test methods (excluding assertion messages) and compare the generated messages against the original ones. Additionally, we explore whether providing additional context—such as test descriptions—enhances the quality of the generated messages.

This paper makes the following contributions:
\begin{enumerate}

\item Empirical Evaluation – We assess the effectiveness of LLMs in generating assertion messages and compare them with human-written messages.

\item Impact of Contextual Information – We investigate whether providing additional information improves message quality.    

\item Providing a comprehensive replication dataset to support reproducibility and facilitate further research\footnote{\url{https://doi.org/10.5281/zenodo.15293133}}.
\end{enumerate}

The remainder of this paper is organized as follows: Section~\ref{sec:RelatedWork} reviews related work on the automated generation of assertion statements. Section~\ref{sec:ResearchMethod} outlines the methodology and experimental design.  Section~\ref{sec:StudyResults} presents the results and key findings derived from our evaluation. Section~\ref{sec:Study_Implication} discusses the broader implications of the findings and provides additional insights. Section~\ref{sec:ThreatsToValidity} outlines the potential limitations and threats to the validity of our study. Finally, Section~\ref{sec:Conclusion} concludes the paper and highlights directions for future work.

%% file: Sections/RelatedWork.tex
\section{Related Work} 
\label{sec:RelatedWork}

This section reviews studies on automated assertion generation in test code, focusing on the role of assertion messages, assertion accuracy, and the application of language models.

Zamprogno et al.~\cite{zamprogno2022dynamic} proposed \textit{AutoAssert}, a tool designed to assist developers in creating assertion statements by capturing runtime values of selected variables during testing. Their evaluation showed that 84.5\% of the generated assertions aligned with developer intent. However, \textit{AutoAssert} exhibits clear limitations—it lacks support for generating informative, custom assertion messages and is restricted to specific assertion types.

He et al.~\cite{he2024empirical} explored deep learning techniques for automating assertion generation, emphasizing the significance of providing contextual details about the Method Under Test (MUT). They demonstrated that including additional context, particularly focal methods, significantly improves assertion quality. However, their approach still struggles to consistently generate assertions that are logically coherent and semantically meaningful.

Primbs et al.~\cite{primbs2025assert5} introduced \textit{AsserT5}, an approach fine-tuning the CodeT5 model specifically for assertion generation. \textit{AsserT5} achieved substantial improvements over existing methods, reaching up to 59.5\% accuracy and successfully identifying 33 real software bugs. Their study underscored the critical influence of input formatting and the targeted fine-tuning of pretrained code models, which significantly enhanced assertion accuracy and reliability. Despite these advancements, their primary focus remained on correctness rather than clarity or informativeness of assertion messages.

Watson et al.~\cite{watson2020learning} designed \textit{ATLAS}, a deep learning framework that treats assertion generation as a translation task, learning from paired test and target methods to produce complete assertion statements. Although their model showed promising accuracy in predicting assertions similar to developer-written statements, it notably did not address generating informative assertion messages explicitly, potentially limiting its effectiveness in debugging contexts.

Zhang et al.~\cite{zhang2025exploring} conducted a comprehensive empirical study assessing the effectiveness of LLMs in assertion generation. Comparing five open-source models against traditional assertion generation techniques, they found that LLMs, particularly CodeT5, consistently achieved higher accuracy. However, their research explicitly focused on generating correct assertion statements rather than informative assertion messages, thus not addressing a crucial aspect that significantly impacts debugging efficiency.

Existing approaches in assertion generation have prioritized correctness and accuracy of assertions, often neglecting the critical role played by informative assertion messages. Frequently, these generated assertions lack descriptive messages. Our work directly addresses this significant gap by evaluating the capabilities of state-of-the-art LLMs, specifically assessing their ability to generate clear, informative assertion messages that facilitate debugging and enhance test code maintainability.

%% file: Sections/ResearchMethod.tex
\section{Research Method}
\label{sec:ResearchMethod}

\begin{figure*}[h!]
    \centering
\includegraphics[width=1\textwidth]{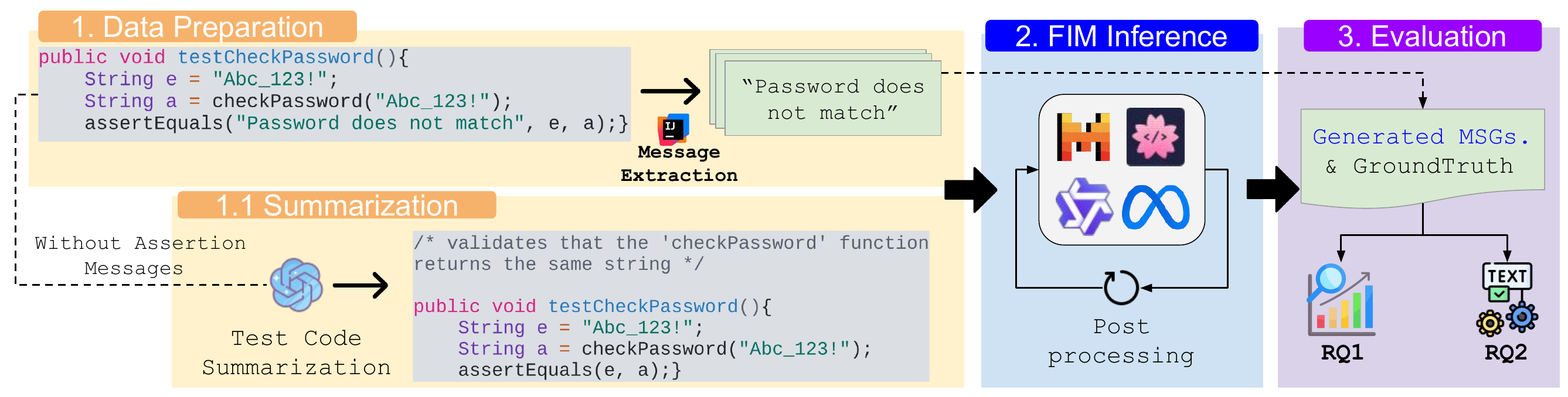}
\vspace*{-10pt}
    \caption{FIM-Based Model Inference Approach.}
    \label{fig:ApporachOverview}
\end{figure*}

Figure~\ref{fig:ApporachOverview} illustrates our methodology. It begins with \textit{Data Preparation}, where we filter Java test methods containing a single assertion from our dataset. For test methods lacking descriptive comments, we generate test-code comments using GPT-4. In the \textit{Inference} phase, we prompt code models (Qwen2.5-Coder-32B, Codestral-22B, CodeLlama-13b-hf, and StarCoder) to generate assertion messages, both without comments and with the generated test comments. Finally, the \textit{Evaluation} phase assesses the performance and structural characteristics of the generated assertion messages using semantic and linguistic methods. 

\vspace*{-3pt}
\subsection{Data Source}
\textbf{Dataset:}
To extract assertion messages, we utilized a publicly available dataset~\cite{takebayashi2023exploratory}\footnote{\url{https://zenodo.org/records/7686565}} consisting of 20 well-engineered Java projects. These projects are known for their high-quality coding standards, clear assertion messages, and have been previously used in multiple studies focused on unit test code analysis and evaluation.

\textbf{Assertion Message Extraction:}
Following the approach described by Takebayashi et al.~\cite{takebayashi2023exploratory}, we used the JetBrains Software Development Kit (SDK)\footnote{\url{https://github.com/JetBrains/intellij-community}} to extract test classes from the selected projects. Specifically, we identified classes containing at least one \texttt{@Test} annotation, indicating actual test methods designed to validate the behavior of a Method Under Test (MUT). Subsequently, we parsed these identified classes to extract individual test methods containing exactly one assertion accompanied by a descriptive assertion message.
The total number of test methods with a single assertion is 508.

\textbf{Filtering:}
To ensure meaningful and reliable ground-truth assertion messages for our evaluation, we applied several preprocessing steps to the extracted dataset. These steps were designed to retain only high-quality assertion messages that clearly provide the reason behind test failures. Specifically, we excluded the following types of assertion messages:

\begin{itemize}
    \item Empty messages, as they do not provide any information.
    \item Non-English messages, since our study and the utilized models are primarily focused on English-language assertions.
    \item Very short or trivial messages, such as single-word assertions or extremely short assertions (less than five characters, e.g., “A”). Such messages have previously been classified as smelly or uninformative assertion messages~\cite{takebayashi2023exploratory}.
    \item Duplicate messages.
\end{itemize}

Applying these filtering criteria resulted in a refined dataset consisting of \textbf{216} test methods with clear, descriptive, and meaningful assertion messages, forming our final ground-truth set for evaluating model-generated assertion messages. In Table~\ref{tab:assertions_types}, we present the distribution of assertion types in our dataset. The most frequent types are assertEquals, assertTrue, and assertFalse. Although this distribution is derived from individual assertion statements, it closely mirrors the distribution of prior work~\cite{takebayashi2023exploratory}.

\textbf{Test Code Summarization—Ablation Comparison RQ1:}
A well-performing model should be able to generate descriptive assertion messages by accurately interpreting the semantic and functional context of the test method. However, providing additional context beyond the raw code—such as comments that describe the test's intent or expected behavior—can significantly enhance the model’s understanding and improve the quality of its generated outputs. Prior work by Primbs et al.~\cite{primbs2025assert5} demonstrated the effectiveness of incorporating method-under-test (MUT) information to improve assertion quality. Similarly, studies in code summarization have shown that enriching source code with code-related information such as data flow and source-code scope leads to better summarization results~\cite{ahmed2024automatic}. 

In our dataset of 216 test methods, only 65 included developer-written comments. However, many of these were low-quality, such as "TODO" comments~\cite{potdar2014exploratory} or references to external resources (e.g., @link to GitHub links or MUT references). After manual filtering, we retained 43 comments that clearly articulated the test’s intent or expected behavior. 
To maintain consistency and fairness in our evaluation, we generated high-quality descriptive comments for the remaining methods using an LLM approach. 
We followed the methodology introduced by Sun et al.~\cite{sun2023automatic}, who showed that GPT-4 (gpt-4-1106-preview) performs well on Java code summarization when prompted with a chain-of-thought (CoT) technique\cite{wei2022chain}, using temperature = 0.1 and top\_p = 0.75. We adopted these exact parameters to generate contextually rich comments for methods lacking meaningful documentation.

\begin{table}[ht]
\centering
\caption{GroundTruth Assertion usage}
\vspace*{-5pt}
\begin{tabular}{l r}
\toprule
\textbf{Assertion Type} & \textbf{Count (\%)} \\
\midrule
assertEquals      & 80 (37.0\%) \\
assertTrue        & 68 (31.5\%) \\
assertFalse       & 31 (14.4\%) \\
assertNotNull     & 11 (5.1\%)  \\
assertThat        & 11 (5.1\%)  \\
assertNull        & 9 (4.2\%)   \\
assertArrayEquals & 2 (0.9\%)   \\
assertNotSame     & 2 (0.9\%)   \\
assertSame        & 2 (0.9\%)   \\
\midrule
\textbf{Total Assertions} & \textbf{216 (100\%)} \\
\bottomrule
\end{tabular}
\label{tab:assertions_types}
\end{table}

\subsection{Code Models and Inference}
\textbf{Objective:} \textit{Given a complete test method with the assertion message removed, the task is to fill in the missing message within the assertion statement.}

\textbf{Models:} Due to the structured nature of this task—where a specific code segment is missing—we opt to use code models trained with a Fill-in-the-Middle (FIM) objective. FIM-capable models are particularly well-suited for such tasks, as they are designed to predict code segments situated between a given prefix and suffix~\cite{bavarian2022efficient}. 
Thus, we evaluate the following base models (i.e., pre-trained models without instruction tuning), selected for their strong performance in recent code generation and infilling benchmarks:
\begin{itemize}
\item Qwen2.5-Coder-32B \cite{hui2024qwen2}\footnote{\url{https://huggingface.co/Qwen/Qwen2.5-Coder-32B}}
\item Codestral-22B-v0.1 \cite{mistral2024codestral}\footnote{\url{https://huggingface.co/mistralai/Codestral-22B-v0.1}}
\item CodeLlama-13B-hf \cite{roziere2023code}\footnote{\url{https://huggingface.co/codellama/CodeLlama-13b-hf}}
\item StarCoder \cite{li2023starcoder}\footnote{\url{https://huggingface.co/bigcode/starcoder}}
\end{itemize}
\noindent
These models have demonstrated strong performance in widely recognized benchmarks such as HumanEvalPlus\footnote{\url{https://evalplus.github.io/leaderboard.html}} \cite{liu2023your} and the HumanEval Infilling Benchmark\footnote{\url{https://github.com/openai/human-eval-infilling?tab=readme-ov-file}} by OpenAI.

\begin{lstlisting}[basicstyle=\small, caption= CodeLlama-13b-hf FIM, label=lst:CodeLlama]
public void testCheckPassword() {
   String e = "Abc_123!";
   String a =checkPassword("Abc_123!");
   //fill_token = tokenizer.fill_token 
   assertEquals(@\textcolor{red}{\texttt{"<FILL\_ME>"}}@, e, a);}
\end{lstlisting}

\begin{lstlisting}[basicstyle=\small, caption=Codestral-22B FIM, label=lst:Codestral]
prefix = """
public void testCheckPassword() {
    String e = "Abc_123!";
    String a = checkPassword("Abc_123!");
    assertEquals("""
suffix = """, e, a);}"""
//from mistral ... import FIMRequest
FIMRequest(prompt=prefix, suffix=suffix)
\end{lstlisting}

\begin{lstlisting}[basicstyle=\small, caption=Qwen2.5-Coder-32B FIM, label=lst:Qwen]
@\textcolor{red}{\texttt{<|fim\_suffix|>}}@
public void testCheckPassword() {
   String e = "Abc_123!";
   String a =checkPassword("Abc_123!");
   assertEquals(@\textcolor{red}{\texttt{<|fim\_suffix|>}}@, e, a);}
   @\textcolor{red}{\texttt{<|fim\_suffix|>}}@
\end{lstlisting}

\noindent
\textbf{Fill-in-the-Middle (FIM) Inference:}
To leverage the infilling capabilities of each model, we format the input test methods using the native FIM prompting structure specific to each model. The missing assertion message is replaced with the model’s designated placeholder or infill token, enabling it to generate code between a provided prefix and suffix.

For example, in Listing~\ref{lst:CodeLlama} (line 5), we demonstrate FIM usage with CodeLlama-13b-hf, where the assertion message is replaced using the token obtained via \texttt{tokenizer.fill\_token}  (i.e., \texttt{<Fill\_ME>}). In Listing~\ref{lst:Codestral}, lines 1 and 6 represent the prefix and suffix segments used with Codestral-22B, while line 8 shows how these components are composed for inference. Listing~\ref{lst:Qwen} (lines 1, 5, and 6) illustrates FIM formatting for Qwen2.5-Coder-32B, following the recommended structure provided by its official documentation\footnote{\url{https://github.com/QwenLM/Qwen2.5-Coder?tab=readme-ov-file}}. StarCoder adopts the same FIM convention as Qwen2.5-Coder-32B, utilizing \texttt{<fim-prefix>}, \texttt{<fim-suffix>}, and \texttt{<fim-middle>} special tokens.

\noindent
For all models, we used the default hyperparameters provided by their respective HuggingFace configurations during inference. This ensures a fair comparison while preserving the intended behavior of each model.

\subsection{Post-processing}
Unlike instruct models—where explicit prompt instructions and illustrative examples of the desired output format can make the output reliable—FIM models inherently rely on special native tags that mark positions for content filling. This reliance introduces two key limitations in their generated outputs.

The first limitation is that FIM models occasionally generate extra tokens beyond the intended assertion messages, such as partial or complete code syntax. A common behavior observed during our evaluation is that the models attempt to generate an entire assertion statement rather than simply providing the assertion message. For example, the model might produce an output like:\texttt{.."CacheLoader should not be closed", loader.closed);}, combining the assertion message with unintended code fragments. To address this issue, we employ a trimming strategy to isolate and extract only the relevant assertion message portion (e.g., extracting \texttt{{"CacheLoader should not be closed"}}). Most models did not require trimming, except for a few cases observed in CodeLlama and Qwen2.5-Coder.

The second limitation arises when the models fail to generate any content at all for a given test method, resulting in empty outputs. To mitigate this, we apply a retry mechanism: whenever a model initially produces an empty assertion message, we resubmit the prompt to ensure the model attempts to generate the missing content.  

Without additional test comments, Qwen2.5-Coder generated 15 empty messages, CodeLlama produced 8, StarCoder produced 3 instances, and Codestral-22B produced none. Incorporating descriptive comments eliminated empty outputs for StarCoder and reduced Qwen’s to 13, while CodeLlama remained at 8 and Codestral continued to produce none. These results highlight Codestral’s consistent reliability in producing non-empty outputs. Applying one or more retries successfully filled every remaining gap, and increased the reliability of the model-generated assertion messages.

\begin{table*}[ht]
\centering
\caption{Performance of LLMs in Generating Assertion Messages (Average Ground truth LLM-Eval Score = 3.24/5). All n-gram overlap metrics scores are measured using the method described in~\cite{zhang2020retrieval, sun2024source}.
$\uparrow$ indicates improvement with additional comments. Bold cells highlight highest scores per metric.}
\vspace*{-8pt}
\resizebox{\textwidth}{!}{
\begin{tabular}{lcccccccccc}
\toprule
 & \multicolumn{5}{c}{\textbf{Test Method Only}} & \multicolumn{5}{c}{\textbf{Test Method with Comments}} \\
\cmidrule(lr){2-6}\cmidrule(lr){7-11}
\textbf{Model} & BLEU & ROUGE-L & METEOR & BERTScore-F1 & LLM-Eval & BLEU & ROUGE-L & METEOR & BERTScore-F1 & LLM-Eval \\
\midrule
CodeLlama-13b-hf & 10.07 & 25.64 & 20.84 & 86.16 & 2.42 & 12.26 $\uparrow$ & 29.71 $\uparrow$ & 25.02 $\uparrow$ & 86.99 $\uparrow$ & 2.63 $\uparrow$ \\
Qwen2.5-Coder-32B & \textbf{15.10} & \textbf{34.60} & \textbf{27.56} & \textbf{87.83} & 2.53 & \textbf{16.17} $\uparrow$ & \textbf{35.48} $\uparrow$ & 30.22 $\uparrow$ & \textbf{88.40} $\uparrow$ & 2.73 $\uparrow$ \\
Codestral-22B & 14.19 & 33.29 & 27.03 & 87.72 & \textbf{2.76} & 15.04 $\uparrow$ & 34.63 $\uparrow$ & \textbf{30.32} $\uparrow$ & 88.02 $\uparrow$ & \textbf{2.97} $\uparrow$ \\
StarCoder & 11.36 & 29.29 & 21.98 & 86.91 & 2.54 & 13.79 $\uparrow$ & 32.48 $\uparrow$ & 27.47 $\uparrow$ & 87.97 $\uparrow$ & 2.83 $\uparrow$ \\
\bottomrule
\end{tabular}}
\label{tab:results_rq1}
\vspace*{5pt}
\end{table*}

\begin{table*}[ht]
\centering
\caption{Distribution of Generated Assertion Message Structures (\%).  Bold cells highlight represent the structure closest to human-written assertion messages (Ground Truth).}
\vspace*{-8pt}
\resizebox{\textwidth}{!}{
\begin{tabular}{lcccccccc}
\toprule
& \multicolumn{4}{c}{\textbf{Test Method Only}} & \multicolumn{4}{c}{\textbf{Test Method with Comments}} \\
\cmidrule(lr){2-5}\cmidrule(lr){6-9}
\textbf{Model} & \textbf{Identifier} & \textbf{String Literal} & \textbf{Combination} & \textbf{Other} & \textbf{Identifier} & \textbf{String Literal} & \textbf{Combination} & \textbf{Other} \\
\midrule
Human (Ground Truth) & 0 & 98.15 & 1.85 & 0 & - & - & - & - \\
CodeLlama-13b-hf & 0.46 & 84.72 & 7.88 & 6.94 & 0 & 90.28 $\uparrow$ & 5.09 & 4.63 $\downarrow$ \\
Qwen2.5-Coder-32B & 0 & 88.89 & 4.17 & 6.94 & 0 & 89.30 $\uparrow$ & 4.65 & 6.05 $\downarrow$ \\
Codestral-22B & 0.93 & \textbf{92.59} & 6.48 & 0  & 0  & 91.67 $\downarrow$ & 8.33 $\uparrow$  & 0 \\
StarCoder & 0.93 & \textbf{91.20} & 5.56 & 2.31 & 0.46 & \textbf{95.37} $\uparrow$ & 4.17 & 0 $\downarrow$ \\
\bottomrule
\end{tabular}}
\label{tab:msg_structure}
\end{table*}

\subsection{Evaluation}
\textbf{Performance:} Assertion message generation is conceptually similar to code summarization~\cite{lu2021codexglue}, where the input is a code snippet and the output is a concise textual description. To evaluate the performance of LLMs in generating assertion messages, we adopt a combination of text similarity and semantic similarity metrics widely used in code summarization research. Specifically, we use BLEU~\cite{papineni2002bleu}, METEOR~\cite{banerjee2005meteor}, and ROUGE-L~\cite{lin2004rouge} to measure n-gram overlap between generated and reference messages. To address the limitations of purely lexical metrics~\cite{haque2022semantic}, we also employ BERTScore~\cite{zhang2019bertscore} to capture deeper semantic similarity. Moreover, recent studies~\cite{sun2023automatic, su2024distilled} have shown that LLM-generated text can often surpass human-written references in quality. This raises concerns about the suitability of relying solely on reference-based similarity metrics. To address this, we adopt an LLM-based evaluation to rank the quality of generated messages. This method captures subjective elements such as usefulness and clarity, which are not easily measured via automatic metrics. Following the approach by Sun et al.~\cite{sun2024source}, we designed a structured prompt to guide the evaluation: 




\begin{acmTitledBox}{Prompt updated based on developer expectations~\cite{peruma2024rationale}}
\noindent 
Here is a Java test method and corresponding assertion messages. Please rate each assertion message on a scale from 1 to 5, where a higher score indicates better quality.
A good assertion message should:

\begin{enumerate}
\item Clearly describe the reason for the assertion failure.
\item Explicitly mention expected versus actual outcomes when relevant.
\item Be concise, informative, and easy to understand.
\item Help developers quickly identify the source of the problem during debugging.
\end{enumerate}

\noindent 
Your answer should follow the format ...
\end{acmTitledBox}

The original evaluation prompt proposed by~\cite{sun2024source} was designed to assess the usefulness of generated source code comments. In our study, we modified this prompt to better suit the context of assertion message generation by aligning it with developer expectations and their understanding of what makes an assertion message clear, helpful, and high-quality~\cite{peruma2024rationale}.

\textbf{Structure and Linguistics:} Unlike traditional code summarization, which typically produces full natural language sentences, assertion messages exhibit more diverse structures~\cite{takebayashi2023exploratory}. They may consist solely of identifiers, plain string literals, or a combination of both. To capture this structural diversity, we extend our evaluation by categorizing the composition of generated messages. However, manual pattern labeling is time-consuming and does not scale well. Fortunately, recent work has shown that LLMs can achieve classification accuracy comparable to manual efforts in various software engineering tasks. For example, Tafreshipour et al.~\cite{tafreshipour2024prompting} used LLaMA3-70B to classify code commits messages into maintenance categories, while Pister et al.~\cite{pister2024promptset} used GPT-4 (gpt-4-1106-preview) to categorize developer-written prompts into multiple types. 
Inspired by these findings, we design a few-shot classification prompt~\cite{wang2020generalizing} to categorize assertion messages into four structural patterns: \texttt{{Identifier, String Literal, Combination, and Other}}. The prompt includes multiple labeled \textit{n}-examples for each category to guide the LLM’s reasoning process. 

Additionally, assertion messages often follow common n-gram patterns, particularly shaped by the type of assertion used. For instance, messages accompanying \texttt{assertEquals} frequently include phrases such as \_“should be”\_ or \_“does not”\_. To evaluate the linguistic alignment between LLM-generated and developer-written messages, we conduct an n-gram analysis of the outputs. Specifically, we extract and compare the top three most frequent bigrams and trigrams from both LLM-generated and reference messages.

%% file: Sections/StudyResults.tex
\vspace*{-5pt}
\section{Results}
\label{sec:StudyResults}

This section outlines the models performance using traditional text similarity metrics, semantic similarity scores, and human-aligned LLM-based evaluation. Additionally, we analyze the structural alignment of the generated assertion messages, including their composition types—Identifier, String Literal, and Combination—as well as n-gram patterns to assess consistency with developer-written messages.
In Figure~\ref{fig:assertion_fim_examples}, we illustrate concrete examples (Listings~\ref{lst:StarCoderExample} through~\ref{lst:QwenExample}) of assertion messages generated by each evaluated model compared to their corresponding human-written messages.

\begin{figure*}[t]
\centering
\begin{minipage}[t]{0.48\textwidth}

\begin{lstlisting}[basicstyle=\small, caption=Example of StarCoder with comment, label=lst:StarCoderExample]
/* Verifies that the contract XML resource
   is available on the classpath. */
public void testContractResourceOnClasspath(){
    URL url = this.getClass().getClassLoader()..
    assertNotNull(url);}
\end{lstlisting}
\textbf{GroundTruth:} "could not find contract resource"\\
\textbf{Generated Msg:} "Contract XML resource not found on classpath"

\end{minipage}
\hfill
\begin{minipage}[t]{0.48\textwidth}

\begin{lstlisting}[basicstyle=\small, caption=Example of CodeLlama, label=lst:CodeLlamaExample]
public void changeAfterBeforeCommit() {
    dirty.afterModification();
    dirty.afterCommit();
    assertThat(dirty.beforeCommit(), is(true));}
\end{lstlisting}
\textbf{GroundTruth:} "Changes after beforeCommit should be dirty after afterCommit"\\
\textbf{Generated Msg:} dirty.beforeCommit()

\end{minipage}

\vspace{2em}

\begin{minipage}[t]{0.48\textwidth}

\begin{lstlisting}[basicstyle=\small, caption=Example of Codestral-22B, label=lst:codestralExample]
public void within() {
    Filter resultFilter = CompilerUtil.parseFilter
    ....
    ....
    assertTrue(resultFilter instanceof Within);}
\end{lstlisting}
\textbf{GroundTruth:} "Within was expected"\\
\textbf{Generated Msg:} "Expected a Within filter but got " + resultFilter.getClass().getName()"

\end{minipage}
\hfill
\begin{minipage}[t]{0.48\textwidth}

\begin{lstlisting}[basicstyle=\small, caption=Example Of Qwen2.5-Coder, label=lst:QwenExample]
public void testSetHead_whenFailed() {
    int maxCapacity = nextPowerOfTwo(16);
    int itemCount = 40;
    ....
    for (int i = 1; i <= itemCount; i++) {
        buffer.add(new TestSequenced(i));}
    long newSequence = 3;
    buffer.setHead(newSequence);
    assertEquals(maxCapacity, buffer.size());}
\end{lstlisting}
\textbf{GroundTruth:} "buffer size should not be affected"\\
\textbf{Generated Msg:} "Buffer size should not change"

\end{minipage}
\caption{Examples of Assertion Message Generation}
\label{fig:assertion_fim_examples}
\end{figure*}

\subsection{RQ1: Performance}
We first investigated how effectively LLMs generate assertion messages by comparing their outputs to the ground truth. Table~\ref{tab:results_rq1} summarizes the evaluation results for each model based on various text and semantic similarity metrics (BLEU, ROUGE-L, METEOR, and BERTScore-F1) and the average human-like evaluation scores provided by an LLM evaluator (LLM-Eval).

First, all models performed below the ground truth level (LLM-Eval = 3.24), indicating room for improvement in generating meaningful assertion messages comparable to those written manually. Among the evaluated models, Codestral-22B consistently received the highest human-like evaluations, scoring 2.76 out of 5 without additional context and improving significantly to 2.97 when additional comments were provided. This suggests that Codestral’s generated messages are perceived as clearer and more useful based on the developer preferences~\cite{peruma2024rationale} compared to those generated by other models.

Second, based on the similarity metrics, Qwen2.5-Coder demonstrated the strongest overall performance. Notably, its BERTScore-F1—a metric that captures deeper semantic similarity—reached 87.83\% using only the test method and increased to 88.40\% when additional contextual information was provided. These results suggest that Qwen2.5-Coder generates assertion messages that are most semantically aligned with the original intent expressed by developers.

Third, CodeLlama showed the weakest performance across all evaluated metrics, both in the test-method-only and context-enhanced settings. Although the model exhibited modest improvements when additional contextual information was provided, its scores consistently lagged behind those of the other models.

Finally, all models demonstrated noticeable improvements with the addition of contextual information in the form of test comments. This positive impact is evident across all evaluation metrics and is especially pronounced in the LLM-Eval scores—for instance, Codestral improved from 2.76 to 2.97, and StarCoder from 2.54 to 2.83. These results support our ablation study findings, confirming that incorporating additional semantic context, such as developer-written comments, enables models to generate assertion messages that are more aligned, clear, and meaningful.

\begin{acmbox}
\noindent 
\textbf{Summary of Performance:} While current LLMs show promising potential in generating informative assertion messages, they still fall short of fully matching the clarity and semantic precision of developer-written messages. Nevertheless, incorporating contextual enhancements—such as descriptive test comments—has been shown to significantly improve their performance. 
\end{acmbox}


\subsection{RQ2: Structure and Linguistics}
This section presents an analysis of the structural and linguistic patterns observed in assertion messages generated by LLMs. Specifically, we examine the types of assertion messages constructs (identifiers, string literals, combinations)  compared  to human-written assertion messages. Additionally, we perform an n-gram analysis to identify common linguistic patterns.

\subsubsection{Assertion Message Structure Analysis}
To better assess the quality and structural alignment of the assertion messages generated by LLMs, we categorized each message into one of four composition-based types (See Section~\ref{sec:ResearchMethod}.4):

\begin{itemize} \item \textbf{Identifier}: Consists solely of code identifiers or expressions (e.g., \texttt{instances.isEmpty()}). \item \textbf{String Literal}: Purely textual, human-readable explanations (e.g., \texttt{``Set should not contain all elements''}). \item \textbf{Combination}: A mixture of string literals and identifiers, often concatenated (e.g., \texttt{``Item mismatch for sequence: '' + e.getKey()}). \item \textbf{Other}: Messages that lack meaningful structure or semantic clarity (e.g., symbols such as \texttt{\textbackslash\textbackslash}). \end{itemize}

Table~\ref{tab:msg_structure} summarizes the distribution of assertion message types generated by each LLM model, compared to the distribution of ground truth.

First, the ground truth messages written by developers show a strong preference for the \textit{String Literal} structure, accounting for 98.15\% of all messages. This highlights a developer tendency toward clear, descriptive, and textual assertion explanations. Only a small fraction (1.85\%) of the messages used a combination of text and code identifiers. These structural patterns and developer preferences are consistent with findings from prior studies~\cite{peruma2024rationale,takebayashi2023exploratory}

Second, all evaluated LLMs predominantly generated \textit{String Literal} assertion messages, reflecting strong alignment with human-written message structures. Codestral-22B (92.59\%) and StarCoder (91.20\%) achieved the highest structural alignment in the test-method-only setting, which further improved when additional contextual information (i.e., test comments) was provided—StarCoder to 95.37\%. Interestingly, Codestral-22B exhibited a slightly different behavior: while test comments effectively reduced the generation of \textit{Identifier} to zero, they also led to a minor decline in pure \textit{String Literal} messages. Instead, the model favored the \textit{Combination} structure, blending textual descriptions with code identifiers.

Third, while some messages fell into the \textit{Combination} and \textit{Other} categories, these less desirable formats were significantly reduced with the inclusion of contextual comments. For example, CodeLlama-13B-hf reduced its generation of \textit{Other}-type messages from 6.94\% to 4.63\% when comments were introduced.

Finally, the near absence of \textit{Identifier}-only messages across all models suggests that LLMs correctly infer the need for natural language explanations in assertion messages, rather than relying solely on code-based identifiers.

\begin{acmbox}
\noindent 
\textbf{Summary of Message Structure:} The structural analysis shows that LLMs largely mimic the compositional patterns of developer-written assertion messages, especially when aided by contextual information. These findings reinforce the value of incorporating descriptive test comments to guide models toward producing clearer and more semantically aligned assertion messages.
\end{acmbox}


\begin{table*}[htbp]
\centering
\caption{Top-3 Most Frequent Bigrams and Trigrams in Assertion Messages (Test Method Only)}
\vspace*{-8pt}
\resizebox{\textwidth}{!}{%
\begin{tabular}{llccc|ccc}
\toprule
 & & \multicolumn{3}{c|}{\textbf{Bigrams}} & \multicolumn{3}{c}{\textbf{Trigrams}} \\
\cmidrule(lr){3-5} \cmidrule(lr){6-8}
\textbf{Source} & \textbf{Assertion Type} & 1st & 2nd & 3rd & 1st & 2nd & 3rd \\
\midrule
Human & assertEquals
& should be & should not & number of
& should not be & should be equal & be equal to \\
& assertTrue
& should be & should have & was expected
& should have been & configuration string should & string should start \\
& assertFalse
& should not & should be & not be
& should not be & not be closed & to no be \\
\midrule
CodeLlama-13b-hf & assertEquals
& should be & but was & number of
& should be the & be the same & size should be \\
& assertTrue
& should be & should have & point1 2
& should have been & an instance of & config should start \\
& assertFalse
& should not & not be & be empty
& should not be & not be empty & should be false \\
\midrule
Codestral-22B & assertEquals
& should be & number of & should not
& should be equal & be equal to & should be empty \\
& assertTrue
& should be & should have & an instance
& an instance of & should be empty & should be an \\
& assertFalse
& should not & not be & be empty
& should not be & not be empty & file should not \\
\midrule
Qwen2.5-Coder-32B & assertEquals
& should be & should not & number of
& should not be & size should be & should be empty \\
& assertTrue
& should be & is not & filter is
& filter is not & is not a & filter filter is \\
& assertFalse
& should not & not be & main has
& should not be & main has subscribers & not be empty \\
\midrule
StarCoder & assertEquals
& should be & should not & not be
& should not be & should be equal & should not change \\
& assertTrue
& should be & filter expected & be empty
& should be empty & should be a & script didnt run \\
& assertFalse
& should not & not be & should be
& should not be & not be empty & file should not \\
\bottomrule
\end{tabular}}
\vspace*{5pt}
\label{tab:ngrams-only}
\end{table*}

\begin{table*}[htbp]
\centering
\caption{Top-3 Most Frequent Bigrams and Trigrams in Assertion Messages (Test Method with Comments)}
\vspace*{-8pt}
\resizebox{\textwidth}{!}{%
\begin{tabular}{llccc|ccc}
\toprule
 & & \multicolumn{3}{c|}{\textbf{Bigrams}} & \multicolumn{3}{c}{\textbf{Trigrams}} \\
\cmidrule(lr){3-5} \cmidrule(lr){6-8}
\textbf{Source} & \textbf{Assertion Type} & 1st & 2nd & 3rd & 1st & 2nd & 3rd \\
\midrule
Human & assertEquals
& should be & should not & number of
& should not be & should be equal & be equal to \\
& assertTrue
& should be & should have & was expected
& should have been & configuration string should & string should start \\
& assertFalse
& should not & should be & not be
& should not be & not be closed & to no be \\
\midrule
CodeLlama-13b-hf & assertEquals
& should be & to be & be equal
& should be equal & be equal to & size should be \\
& assertTrue
& should be & an instance & instance of
& an instance of & should start with & should be empty \\
& assertFalse
& should not & not be & be empty
& should not be & not be empty & set should not \\
\midrule
Codestral-22B & assertEquals
& should be & should not & number of
& should not be & should be empty & are not equal \\
& assertTrue
& should be & be empty & did not
& should start with & should be empty & be an instance \\
& assertFalse
& should not & not be & be empty
& should not be & not be empty & be empty after \\
\midrule
Qwen2.5-Coder-32B & assertEquals
& should be & should not & not be
& should not be & should be empty & size should be \\
& assertTrue
& should be & is not & filter is
& filter is not & is not a & should be empty \\
& assertFalse
& should not & not be & has subscribers
& should not be & not be empty & not be closed \\
\midrule
StarCoder & assertEquals
& should be & number of & should not
& incorrect number of & should be equal & be equal to \\
& assertTrue
& should be & not found & expected a
& not found in & should have been & an instance of \\
& assertFalse
& should not & not be & be empty
& should not be & not be empty & file should not \\
\bottomrule
\end{tabular}}
\label{tab:ngrams-comments}
\end{table*}

\subsubsection{N-Gram Analysis of Assertion Messages}
To gain deeper insights into the linguistic patterns and structural alignment of generated assertion messages, we conducted a detailed n-gram analysis focusing on the most frequent bigrams (two-word sequences) and trigrams (three-word sequences) produced by each LLM. Tables~\ref{tab:ngrams-only} and~\ref{tab:ngrams-comments} summarize the top three most frequently occurring patterns, categorized by assertion types (\texttt{assertEquals}, \texttt{assertTrue}, and \texttt{assertFalse}) as they most occurred in the ground truth Table~\ref{tab:assertions_types}.

In Table~\ref{tab:ngrams-only} (Test Method Only), all evaluated models closely replicated the most frequent developer-written bigrams. Specifically, models consistently generated the bigram \textit{"should be"} for \texttt{assertEquals} and \texttt{assertTrue}, and \textit{"should not"} for \texttt{assertFalse}, closely matching human preferences.

For the most frequent trigrams, developers commonly used the phrase \textit{"should not be"} for both \texttt{assertEquals} and \texttt{assertFalse}. Only Qwen2.5-Coder and StarCoder correctly reproduced \textit{"should not be"} for \texttt{assertEquals}, while all models matched this trigram for \texttt{assertFalse}. Additionally, developers frequently used \textit{"should have been"} for \texttt{assertTrue}, which was accurately reproduced only by CodeLlama. Other models introduced less meaningful alternatives, such as Qwen2.5-Coder’s \textit{"filter is not"}.

The second-most frequent bigrams revealed additional insights. Developers used \textit{"should not"} for \texttt{assertEquals}, which was matched by Qwen2.5-Coder and StarCoder. For \texttt{assertTrue}, the developer-preferred bigram \textit{"should have"} was correctly reproduced by CodeLlama and Codestral-22B. Interestingly, while developers used \textit{"should be"} as the second-most frequent bigram for \texttt{assertFalse}, no model replicated it. Instead, all models produced the third-most frequent developer bigram, \textit{"not be"}.

Table~\ref{tab:ngrams-comments} (Test Method with Comments) shows the improvements achieved when additional context was provided. Models retained strong alignment for the most frequent bigrams. Notably, Qwen2.5-Coder-32B maintained its alignment, continuing to generate \textit{"should not"} for \texttt{assertEquals}, while Codestral-22B improved from generating \textit{"number of"} (without comments) to \textit{"should not"} (with comments)—which matches the developer bigrams.

Certain models, particularly CodeLlama and Qwen2.5-Coder, introduced problematic or overly context-specific phrases. For instance, CodeLlama generated the unclear bigram \textit{"point1 2"} in Table~\ref{tab:ngrams-only}, while Qwen2.5-Coder produced the less meaningful trigram \textit{"filter filter is"}, revealing difficulties in capturing semantic intent. However, the addition of contextual comments (Table~\ref{tab:ngrams-comments}) improved the linguistic clarity of these outputs. CodeLlama eliminated \textit{"point1 2"} and instead produced clearer bigrams such as \textit{"instance of"}. Similarly, Qwen2.5-Coder improved from \textit{"filter filter is"} to \textit{"should be empty"}.

\begin{acmbox}
\noindent 
\textbf{Summary of N-gram Analysis:}
The analysis demonstrates that LLMs effectively replicate common human-written linguistic patterns, especially for frequent bigrams. Introducing contextual information moderately improves N-gram precision.
\end{acmbox}


%% file: Sections/Study_Implication.tex
\section{Additional Discussions and Implications} 
\label{sec:Study_Implication}

\textbf{Models performance:} Although LLMs are promising tools for automatically generating assertion messages, our evaluation indicates they have not yet reached the human-written messages, particularly in terms of the human baseline (LLM-Eval score of 3.24), even when augmented with test comments.
Each model showed distinct strengths and weaknesses in assertion message generation. Qwen2.5-Coder excelled in semantic accuracy as indicated by high BERTScore-F1 values, suggesting superior comprehension of test intent compared to others. Codestral-22B achieved the best performance according to human-like evaluation (LLM-Eval), indicating that it produces the most developer-friendly, comprehensible messages. CodeLlama-13b-hf, conversely, consistently lagged behind in both semantic alignment and clarity, suggesting that further fine-tuning on assertion-message-specific datasets may significantly improve its performance. Therefore, selecting models aligned with specific quality objectives is essential when adopting LLMs for assertion message generation—whether the goal is maximizing semantic precision (e.g., Qwen2.5-Coder-32B) or enhancing human-readable clarity and consistency (e.g., Codestral-22B).

\textbf{Contextual Information--Comments:} The ablation study (in RQ1) clearly demonstrates that enriching test methods with additional semantic context significantly enhances the quality of generated assertion messages. Specifically, the inclusion of test comments notably improved human-like similarity (e.g., Codestral-22B improved from an LLM-Eval score of 2.76 to 2.97, and similar improvements were seen across other metrics and models). This reinforces the importance of well-written test documentation not only aids human developers but also directly supports more effective LLM-driven test automation practices.

\textbf{Structural Alignment and Linguistic Patterns:}
The analysis of assertion message structures and linguistic patterns (i.e., bigrams and trigrams) reveals that LLMs are generally effective at replicating common linguistic conventions used by developers—particularly frequent phrases such as \textit{''should be''} and \textit{''should not be''}. While the evaluated models often adhered to these human-like patterns, structural inconsistencies were occasionally observed, including messages dominated by identifiers or code-centric artifacts (categorized as \textit{Other}). Our results indicate that models such as CodeLlama and Qwen2.5-Coder were less reliable in avoiding such code-based structures, especially compared to models like Codestral and StarCoder, which produced more consistently human-readable outputs. These findings underscore the need for larger and more diverse datasets of high-quality assertion messages—not only to improve benchmarking and evaluation—but also to support fine-tuning efforts aimed at reducing code-like artifacts and improving linguistic alignment with developer expectations.

\textbf{Evaluation Metrics for Assertion Message:}
Our study utilized standard textual and semantic similarity metrics commonly adopted in code summarization tasks, including BLEU, ROUGE-L, METEOR, and BERTScore-F1. However, the nature of assertion messages differs significantly from conventional code summaries, which are typically longer and written as natural language sentences. Assertion messages, by contrast, are often shorter, more concise, and structurally diverse. As highlighted in our structural analysis (section~\ref{sec:ResearchMethod}.4 and section ~\ref{sec:StudyResults}-RQ2), these messages may appear as short textual descriptions, code-based identifiers, or combinations of both.
This inherent structural variability introduces challenges in applying standard evaluation metrics. Lexical similarity metrics such as BLEU, ROUGE-L, and METEOR rely heavily on word-level overlap, which inherently favors messages written in fluent natural language. As a result, they may undervalue assertion messages that are semantically correct but expressed using concise code-like forms or identifier-based phrasing. Although BERTScore-F1 is designed to capture semantic similarity at the embedding level, it is primarily trained on natural language text~\cite{zhang2019bertscore} and may not fully capture the details of shorter text or code-centric assertion messages. Consequently, while these metrics offer useful approximations of quality, their effectiveness may be limited when evaluating assertion messages.

%% file: Sections/ThreatsToValidity.tex
\section{Threats to Validity}
\label{sec:ThreatsToValidity}
\textbf{External Validity:}
One concern regarding our study pertains to the generalizability of the results. Our investigation focused exclusively on Java test methods utilizing single assertion messages within JUnit4 frameworks from  20 well-engineered projects. Consequently, the observed model behaviors and insights may not directly generalize to other Java testing frameworks, such as JUnit5 or TestNG, nor to different programming languages, such as Python. Additionally, our analysis was restricted specifically to test methods containing exactly one assertion. Thus, the applicability of our findings to more complex test methods containing multiple assertions remains unexplored, representing an important direction for future research.

\vspace*{3pt}
\textbf{Internal Validity:}
Our evaluation of assertion message quality (LLM-Eval), structural classification, and test comments relied on GPT-4, potentially introducing evaluator bias or inaccuracies from automated classification. To mitigate this, we conducted manual reviews of LLM outputs and closely followed validated methods from prior work~\cite{tafreshipour2024prompting, pister2024promptset, sun2024source}. 
Moreover, we used default inference hyperparameters provided by each model’s HuggingFace repository, without task-specific fine-tuning. Although this ensured fairness in comparing models, it may not represent each model's optimal performance. Lastly, we exclusively evaluated base-model LLMs trained with the Fill-in-the-Middle (FIM) objective, excluding instruction-tuned or commercial models (e.g., OpenAI’s GPT series). While our choice established  a clear baseline of the models performance--as this study represents the first systematic investigation into LLM-generated assertion messages, future research could explore hyperparameter optimization or instruction-tuned models explicitly tailored for assertion message generation, potentially uncovering further performance improvements.

%% file: Sections/Conclusion.tex
\section{Conclusions}
\label{sec:Conclusion}
In this study, we systematically evaluated four state-of-the-art Fill-in-the-Middle (FIM) large language models—Qwen2.5-Coder-32B, Codestral-22B, CodeLlama-13b-hf, and StarCoder—for their ability to generate informative assertion messages in Java test methods. Our results demonstrate that while current LLMs, notably Codestral-22B, achieve promising results, they still fall short of matching human-written assertion messages in terms of clarity and semantic accuracy. Our ablation study further highlights the significant benefit of including contextual information, such as descriptive test comments, which considerably improves the semantic precision and readability of the generated messages. Structural and linguistic analyses confirm that LLMs effectively replicate common developer patterns but reveal challenges related to semantic clarity in concise, mixed-format messages. Lastly, our findings suggest that existing textual evaluation metrics have limitations when applied to short, structurally diverse assertion messages. Our work establishes foundational insights and benchmarks to guide future research toward automated and context-aware assertion message generation, ultimately aiming to enhance test code quality and maintainability.

%% file: main.bbl

\begin{thebibliography}{40}


\ifx \showCODEN    \undefined \def \showCODEN     #1{\unskip}     \fi
\ifx \showISBNx    \undefined \def \showISBNx     #1{\unskip}     \fi
\ifx \showISBNxiii \undefined \def \showISBNxiii  #1{\unskip}     \fi
\ifx \showISSN     \undefined \def \showISSN      #1{\unskip}     \fi
\ifx \showLCCN     \undefined \def \showLCCN      #1{\unskip}     \fi
\ifx \shownote     \undefined \def \shownote      #1{#1}          \fi
\ifx \showarticletitle \undefined \def \showarticletitle #1{#1}   \fi
\ifx \showURL      \undefined \def \showURL       {\relax}        \fi
\providecommand\bibfield[2]{#2}
\providecommand\bibinfo[2]{#2}
\providecommand\natexlab[1]{#1}
\providecommand\showeprint[2][]{arXiv:#2}

\bibitem[Ahmed et~al\mbox{.}(2024)]%
        {ahmed2024automatic}
\bibfield{author}{\bibinfo{person}{Toufique Ahmed}, \bibinfo{person}{Kunal~Suresh Pai}, \bibinfo{person}{Premkumar Devanbu}, {and} \bibinfo{person}{Earl Barr}.} \bibinfo{year}{2024}\natexlab{}.
\newblock \showarticletitle{Automatic semantic augmentation of language model prompts (for code summarization)}. In \bibinfo{booktitle}{\emph{Proceedings of the IEEE/ACM 46th international conference on software engineering}}. \bibinfo{pages}{1--13}.
\newblock


\bibitem[AI(2024)]%
        {mistral2024codestral}
\bibfield{author}{\bibinfo{person}{Mistral AI}.} \bibinfo{year}{2024}\natexlab{}.
\newblock \bibinfo{title}{Introducing Codestral: A State-of-the-Art Code Language Model}.
\newblock \bibinfo{howpublished}{\url{https://mistral.ai/news/codestral-2501}}.
\newblock
\newblock
\shownote{Accessed: 2025-02-25}.


\bibitem[Aljedaani et~al\mbox{.}(2021)]%
        {aljedaani2021test}
\bibfield{author}{\bibinfo{person}{Wajdi Aljedaani}, \bibinfo{person}{Anthony Peruma}, \bibinfo{person}{Ahmed Aljohani}, \bibinfo{person}{Mazen Alotaibi}, \bibinfo{person}{Mohamed~Wiem Mkaouer}, \bibinfo{person}{Ali Ouni}, \bibinfo{person}{Christian~D Newman}, \bibinfo{person}{Abdullatif Ghallab}, {and} \bibinfo{person}{Stephanie Ludi}.} \bibinfo{year}{2021}\natexlab{}.
\newblock \showarticletitle{Test smell detection tools: A systematic mapping study}. In \bibinfo{booktitle}{\emph{Proceedings of the 25th International Conference on Evaluation and Assessment in Software Engineering}}. \bibinfo{pages}{170--180}.
\newblock


\bibitem[Aljohani and Do(2024)]%
        {aljohani2024fine}
\bibfield{author}{\bibinfo{person}{Ahmed Aljohani} {and} \bibinfo{person}{Hyunsook Do}.} \bibinfo{year}{2024}\natexlab{}.
\newblock \showarticletitle{From fine-tuning to output: An empirical investigation of test smells in transformer-based test code generation}. In \bibinfo{booktitle}{\emph{Proceedings of the 39th ACM/SIGAPP Symposium on Applied Computing}}. \bibinfo{pages}{1282--1291}.
\newblock


\bibitem[Banerjee and Lavie(2005)]%
        {banerjee2005meteor}
\bibfield{author}{\bibinfo{person}{Satanjeev Banerjee} {and} \bibinfo{person}{Alon Lavie}.} \bibinfo{year}{2005}\natexlab{}.
\newblock \showarticletitle{METEOR: An automatic metric for MT evaluation with improved correlation with human judgments}. In \bibinfo{booktitle}{\emph{Proceedings of the acl workshop on intrinsic and extrinsic evaluation measures for machine translation and/or summarization}}. \bibinfo{pages}{65--72}.
\newblock


\bibitem[Bavarian et~al\mbox{.}(2022)]%
        {bavarian2022efficient}
\bibfield{author}{\bibinfo{person}{Mohammad Bavarian}, \bibinfo{person}{Heewoo Jun}, \bibinfo{person}{Nikolas Tezak}, \bibinfo{person}{John Schulman}, \bibinfo{person}{Christine McLeavey}, \bibinfo{person}{Jerry Tworek}, {and} \bibinfo{person}{Mark Chen}.} \bibinfo{year}{2022}\natexlab{}.
\newblock \showarticletitle{Efficient training of language models to fill in the middle}.
\newblock \bibinfo{journal}{\emph{arXiv preprint arXiv:2207.14255}} (\bibinfo{year}{2022}).
\newblock


\bibitem[Bavota et~al\mbox{.}(2015)]%
        {bavota2015test}
\bibfield{author}{\bibinfo{person}{Gabriele Bavota}, \bibinfo{person}{Abdallah Qusef}, \bibinfo{person}{Rocco Oliveto}, \bibinfo{person}{Andrea De~Lucia}, {and} \bibinfo{person}{Dave Binkley}.} \bibinfo{year}{2015}\natexlab{}.
\newblock \showarticletitle{Are test smells really harmful? an empirical study}.
\newblock \bibinfo{journal}{\emph{Empirical Software Engineering}}  \bibinfo{volume}{20} (\bibinfo{year}{2015}), \bibinfo{pages}{1052--1094}.
\newblock


\bibitem[Deljouyi et~al\mbox{.}(2024)]%
        {deljouyi2024leveraging}
\bibfield{author}{\bibinfo{person}{Amirhossein Deljouyi}, \bibinfo{person}{Roham Koohestani}, \bibinfo{person}{Maliheh Izadi}, {and} \bibinfo{person}{Andy Zaidman}.} \bibinfo{year}{2024}\natexlab{}.
\newblock \showarticletitle{Leveraging large language models for enhancing the understandability of generated unit tests}.
\newblock \bibinfo{journal}{\emph{arXiv preprint arXiv:2408.11710}} (\bibinfo{year}{2024}).
\newblock


\bibitem[Fraser and Arcuri(2011)]%
        {fraser2011evosuite}
\bibfield{author}{\bibinfo{person}{Gordon Fraser} {and} \bibinfo{person}{Andrea Arcuri}.} \bibinfo{year}{2011}\natexlab{}.
\newblock \showarticletitle{Evosuite: automatic test suite generation for object-oriented software}. In \bibinfo{booktitle}{\emph{Proceedings of the 19th ACM SIGSOFT symposium and the 13th European conference on Foundations of software engineering}}. \bibinfo{pages}{416--419}.
\newblock


\bibitem[Haque et~al\mbox{.}(2022)]%
        {haque2022semantic}
\bibfield{author}{\bibinfo{person}{Sakib Haque}, \bibinfo{person}{Zachary Eberhart}, \bibinfo{person}{Aakash Bansal}, {and} \bibinfo{person}{Collin McMillan}.} \bibinfo{year}{2022}\natexlab{}.
\newblock \showarticletitle{Semantic similarity metrics for evaluating source code summarization}. In \bibinfo{booktitle}{\emph{Proceedings of the 30th IEEE/ACM International Conference on Program Comprehension}}. \bibinfo{pages}{36--47}.
\newblock


\bibitem[He et~al\mbox{.}(2024)]%
        {he2024empirical}
\bibfield{author}{\bibinfo{person}{Yibo He}, \bibinfo{person}{Jiaming Huang}, \bibinfo{person}{Hao Yu}, {and} \bibinfo{person}{Tao Xie}.} \bibinfo{year}{2024}\natexlab{}.
\newblock \showarticletitle{An empirical study on focal methods in deep-learning-based approaches for assertion generation}.
\newblock \bibinfo{journal}{\emph{Proceedings of the ACM on Software Engineering}} \bibinfo{volume}{1}, \bibinfo{number}{FSE} (\bibinfo{year}{2024}), \bibinfo{pages}{1750--1771}.
\newblock


\bibitem[Hui et~al\mbox{.}(2024)]%
        {hui2024qwen2}
\bibfield{author}{\bibinfo{person}{Binyuan Hui}, \bibinfo{person}{Jian Yang}, \bibinfo{person}{Zeyu Cui}, \bibinfo{person}{Jiaxi Yang}, \bibinfo{person}{Dayiheng Liu}, \bibinfo{person}{Lei Zhang}, \bibinfo{person}{Tianyu Liu}, \bibinfo{person}{Jiajun Zhang}, \bibinfo{person}{Bowen Yu}, \bibinfo{person}{Kai Dang}, {et~al\mbox{.}}} \bibinfo{year}{2024}\natexlab{}.
\newblock \showarticletitle{Qwen2. 5-Coder Technical Report}.
\newblock \bibinfo{journal}{\emph{arXiv preprint arXiv:2409.12186}} (\bibinfo{year}{2024}).
\newblock


\bibitem[Khorikov(2020)]%
        {khorikov2020unit}
\bibfield{author}{\bibinfo{person}{Vladimir Khorikov}.} \bibinfo{year}{2020}\natexlab{}.
\newblock \bibinfo{booktitle}{\emph{Unit testing principles, practices, and patterns}}.
\newblock \bibinfo{publisher}{Simon and Schuster}.
\newblock


\bibitem[Li et~al\mbox{.}(2023)]%
        {li2023starcoder}
\bibfield{author}{\bibinfo{person}{Raymond Li}, \bibinfo{person}{Loubna~Ben Allal}, \bibinfo{person}{Yangtian Zi}, \bibinfo{person}{Niklas Muennighoff}, \bibinfo{person}{Denis Kocetkov}, \bibinfo{person}{Chenghao Mou}, \bibinfo{person}{Marc Marone}, \bibinfo{person}{Christopher Akiki}, \bibinfo{person}{Jia Li}, \bibinfo{person}{Jenny Chim}, \bibinfo{person}{Qian Liu}, \bibinfo{person}{Evgenii Zheltonozhskii}, \bibinfo{person}{Terry~Yue Zhuo}, \bibinfo{person}{Thomas Wang}, \bibinfo{person}{Olivier Dehaene}, \bibinfo{person}{Mishig Davaadorj}, \bibinfo{person}{Joel Lamy-Poirier}, \bibinfo{person}{João Monteiro}, \bibinfo{person}{Oleh Shliazhko}, \bibinfo{person}{Nicolas Gontier}, \bibinfo{person}{Nicholas Meade}, \bibinfo{person}{Armel Zebaze}, \bibinfo{person}{Ming-Ho Yee}, \bibinfo{person}{Logesh~Kumar Umapathi}, \bibinfo{person}{Jian Zhu}, \bibinfo{person}{Benjamin Lipkin}, \bibinfo{person}{Muhtasham Oblokulov}, \bibinfo{person}{Zhiruo Wang}, \bibinfo{person}{Rudra Murthy}, \bibinfo{person}{Jason
  Stillerman}, \bibinfo{person}{Siva~Sankalp Patel}, \bibinfo{person}{Dmitry Abulkhanov}, \bibinfo{person}{Marco Zocca}, \bibinfo{person}{Manan Dey}, \bibinfo{person}{Zhihan Zhang}, \bibinfo{person}{Nour Fahmy}, \bibinfo{person}{Urvashi Bhattacharyya}, \bibinfo{person}{Wenhao Yu}, \bibinfo{person}{Swayam Singh}, \bibinfo{person}{Sasha Luccioni}, \bibinfo{person}{Paulo Villegas}, \bibinfo{person}{Maxim Kunakov}, \bibinfo{person}{Fedor Zhdanov}, \bibinfo{person}{Manuel Romero}, \bibinfo{person}{Tony Lee}, \bibinfo{person}{Nadav Timor}, \bibinfo{person}{Jennifer Ding}, \bibinfo{person}{Claire Schlesinger}, \bibinfo{person}{Hailey Schoelkopf}, \bibinfo{person}{Jan Ebert}, \bibinfo{person}{Tri Dao}, \bibinfo{person}{Mayank Mishra}, \bibinfo{person}{Alex Gu}, \bibinfo{person}{Jennifer Robinson}, \bibinfo{person}{Carolyn~Jane Anderson}, \bibinfo{person}{Brendan Dolan-Gavitt}, \bibinfo{person}{Danish Contractor}, \bibinfo{person}{Siva Reddy}, \bibinfo{person}{Daniel Fried}, \bibinfo{person}{Dzmitry Bahdanau},
  \bibinfo{person}{Yacine Jernite}, \bibinfo{person}{Carlos~Muñoz Ferrandis}, \bibinfo{person}{Sean Hughes}, \bibinfo{person}{Thomas Wolf}, \bibinfo{person}{Arjun Guha}, \bibinfo{person}{Leandro von Werra}, {and} \bibinfo{person}{Harm de Vries}.} \bibinfo{year}{2023}\natexlab{}.
\newblock \showarticletitle{StarCoder: may the source be with you!}
\newblock  (\bibinfo{year}{2023}).
\newblock
\showeprint[arxiv]{2305.06161}~[cs.CL]


\bibitem[Lin(2004)]%
        {lin2004rouge}
\bibfield{author}{\bibinfo{person}{Chin-Yew Lin}.} \bibinfo{year}{2004}\natexlab{}.
\newblock \showarticletitle{Rouge: A package for automatic evaluation of summaries}. In \bibinfo{booktitle}{\emph{Text summarization branches out}}. \bibinfo{pages}{74--81}.
\newblock


\bibitem[Liu et~al\mbox{.}(2023)]%
        {liu2023your}
\bibfield{author}{\bibinfo{person}{Jiawei Liu}, \bibinfo{person}{Chunqiu~Steven Xia}, \bibinfo{person}{Yuyao Wang}, {and} \bibinfo{person}{Lingming Zhang}.} \bibinfo{year}{2023}\natexlab{}.
\newblock \showarticletitle{Is your code generated by chatgpt really correct? rigorous evaluation of large language models for code generation}.
\newblock \bibinfo{journal}{\emph{arXiv preprint arXiv:2305.01210}} (\bibinfo{year}{2023}).
\newblock


\bibitem[Lu et~al\mbox{.}(2021)]%
        {lu2021codexglue}
\bibfield{author}{\bibinfo{person}{Shuai Lu}, \bibinfo{person}{Daya Guo}, \bibinfo{person}{Shuo Ren}, \bibinfo{person}{Junjie Huang}, \bibinfo{person}{Alexey Svyatkovskiy}, \bibinfo{person}{Ambrosio Blanco}, \bibinfo{person}{Colin Clement}, \bibinfo{person}{Dawn Drain}, \bibinfo{person}{Daxin Jiang}, \bibinfo{person}{Duyu Tang}, {et~al\mbox{.}}} \bibinfo{year}{2021}\natexlab{}.
\newblock \showarticletitle{Codexglue: A machine learning benchmark dataset for code understanding and generation}.
\newblock \bibinfo{journal}{\emph{arXiv preprint arXiv:2102.04664}} (\bibinfo{year}{2021}).
\newblock


\bibitem[Ou{\'e}draogo et~al\mbox{.}(2024)]%
        {ouedraogo2024test}
\bibfield{author}{\bibinfo{person}{Wendk{\^u}uni~C Ou{\'e}draogo}, \bibinfo{person}{Yinghua Li}, \bibinfo{person}{Kader Kabor{\'e}}, \bibinfo{person}{Xunzhu Tang}, \bibinfo{person}{Anil Koyuncu}, \bibinfo{person}{Jacques Klein}, \bibinfo{person}{David Lo}, {and} \bibinfo{person}{Tegawend{\'e}~F Bissyand{\'e}}.} \bibinfo{year}{2024}\natexlab{}.
\newblock \showarticletitle{Test smells in LLM-Generated Unit Tests}.
\newblock \bibinfo{journal}{\emph{arXiv preprint arXiv:2410.10628}} (\bibinfo{year}{2024}).
\newblock


\bibitem[Panichella et~al\mbox{.}(2022)]%
        {panichella2022test}
\bibfield{author}{\bibinfo{person}{Annibale Panichella}, \bibinfo{person}{Sebastiano Panichella}, \bibinfo{person}{Gordon Fraser}, \bibinfo{person}{Anand~Ashok Sawant}, {and} \bibinfo{person}{Vincent~J Hellendoorn}.} \bibinfo{year}{2022}\natexlab{}.
\newblock \showarticletitle{Test smells 20 years later: detectability, validity, and reliability}.
\newblock \bibinfo{journal}{\emph{Empirical Software Engineering}} \bibinfo{volume}{27}, \bibinfo{number}{7} (\bibinfo{year}{2022}), \bibinfo{pages}{170}.
\newblock


\bibitem[Papineni et~al\mbox{.}(2002)]%
        {papineni2002bleu}
\bibfield{author}{\bibinfo{person}{Kishore Papineni}, \bibinfo{person}{Salim Roukos}, \bibinfo{person}{Todd Ward}, {and} \bibinfo{person}{Wei-Jing Zhu}.} \bibinfo{year}{2002}\natexlab{}.
\newblock \showarticletitle{Bleu: a method for automatic evaluation of machine translation}. In \bibinfo{booktitle}{\emph{Proceedings of the 40th annual meeting of the Association for Computational Linguistics}}. \bibinfo{pages}{311--318}.
\newblock


\bibitem[Peruma et~al\mbox{.}(2024)]%
        {peruma2024rationale}
\bibfield{author}{\bibinfo{person}{Anthony Peruma}, \bibinfo{person}{Taryn Takebayashi}, \bibinfo{person}{Rocky Huang}, \bibinfo{person}{Joseph~Carmelo Averion}, \bibinfo{person}{Veronica Hodapp}, \bibinfo{person}{Christian~D Newman}, {and} \bibinfo{person}{Mohamed~Wiem Mkaouer}.} \bibinfo{year}{2024}\natexlab{}.
\newblock \showarticletitle{On the Rationale and Use of Assertion Messages in Test Code: Insights from Software Practitioners}. In \bibinfo{booktitle}{\emph{2024 IEEE International Conference on Software Maintenance and Evolution (ICSME)}}. IEEE, \bibinfo{pages}{538--549}.
\newblock


\bibitem[Pister et~al\mbox{.}(2024)]%
        {pister2024promptset}
\bibfield{author}{\bibinfo{person}{Kaiser Pister}, \bibinfo{person}{Dhruba~Jyoti Paul}, \bibinfo{person}{Ishan Joshi}, {and} \bibinfo{person}{Patrick Brophy}.} \bibinfo{year}{2024}\natexlab{}.
\newblock \showarticletitle{PromptSet: A Programmer's Prompting Dataset}. In \bibinfo{booktitle}{\emph{Proceedings of the 1st International Workshop on Large Language Models for Code}}. \bibinfo{pages}{62--69}.
\newblock


\bibitem[Potdar and Shihab(2014)]%
        {potdar2014exploratory}
\bibfield{author}{\bibinfo{person}{Aniket Potdar} {and} \bibinfo{person}{Emad Shihab}.} \bibinfo{year}{2014}\natexlab{}.
\newblock \showarticletitle{An exploratory study on self-admitted technical debt}. In \bibinfo{booktitle}{\emph{2014 IEEE International Conference on Software Maintenance and Evolution}}. IEEE, \bibinfo{pages}{91--100}.
\newblock


\bibitem[Primbs et~al\mbox{.}(2025)]%
        {primbs2025assert5}
\bibfield{author}{\bibinfo{person}{Severin Primbs}, \bibinfo{person}{Benedikt Fein}, {and} \bibinfo{person}{Gordon Fraser}.} \bibinfo{year}{2025}\natexlab{}.
\newblock \showarticletitle{AsserT5: Test Assertion Generation Using a Fine-Tuned Code Language Model}.
\newblock \bibinfo{journal}{\emph{arXiv preprint arXiv:2502.02708}} (\bibinfo{year}{2025}).
\newblock


\bibitem[Roziere et~al\mbox{.}(2023)]%
        {roziere2023code}
\bibfield{author}{\bibinfo{person}{Baptiste Roziere}, \bibinfo{person}{Jonas Gehring}, \bibinfo{person}{Fabian Gloeckle}, \bibinfo{person}{Sten Sootla}, \bibinfo{person}{Itai Gat}, \bibinfo{person}{Xiaoqing~Ellen Tan}, \bibinfo{person}{Yossi Adi}, \bibinfo{person}{Jingyu Liu}, \bibinfo{person}{Romain Sauvestre}, \bibinfo{person}{Tal Remez}, {et~al\mbox{.}}} \bibinfo{year}{2023}\natexlab{}.
\newblock \showarticletitle{Code llama: Open foundation models for code}.
\newblock \bibinfo{journal}{\emph{arXiv preprint arXiv:2308.12950}} (\bibinfo{year}{2023}).
\newblock


\bibitem[Siddiq et~al\mbox{.}(2024)]%
        {siddiq2024using}
\bibfield{author}{\bibinfo{person}{Mohammed~Latif Siddiq}, \bibinfo{person}{Joanna~Cecilia Da~Silva~Santos}, \bibinfo{person}{Ridwanul~Hasan Tanvir}, \bibinfo{person}{Noshin Ulfat}, \bibinfo{person}{Fahmid Al~Rifat}, {and} \bibinfo{person}{Vin{\'\i}cius Carvalho~Lopes}.} \bibinfo{year}{2024}\natexlab{}.
\newblock \showarticletitle{Using large language models to generate junit tests: An empirical study}. In \bibinfo{booktitle}{\emph{Proceedings of the 28th International Conference on Evaluation and Assessment in Software Engineering}}. \bibinfo{pages}{313--322}.
\newblock


\bibitem[Su and McMillan(2024)]%
        {su2024distilled}
\bibfield{author}{\bibinfo{person}{Chia-Yi Su} {and} \bibinfo{person}{Collin McMillan}.} \bibinfo{year}{2024}\natexlab{}.
\newblock \showarticletitle{Distilled GPT for source code summarization}.
\newblock \bibinfo{journal}{\emph{Automated Software Engineering}} \bibinfo{volume}{31}, \bibinfo{number}{1} (\bibinfo{year}{2024}), \bibinfo{pages}{22}.
\newblock


\bibitem[Sun et~al\mbox{.}(2023)]%
        {sun2023automatic}
\bibfield{author}{\bibinfo{person}{Weisong Sun}, \bibinfo{person}{Chunrong Fang}, \bibinfo{person}{Yudu You}, \bibinfo{person}{Yun Miao}, \bibinfo{person}{Yi Liu}, \bibinfo{person}{Yuekang Li}, \bibinfo{person}{Gelei Deng}, \bibinfo{person}{Shenghan Huang}, \bibinfo{person}{Yuchen Chen}, \bibinfo{person}{Quanjun Zhang}, {et~al\mbox{.}}} \bibinfo{year}{2023}\natexlab{}.
\newblock \showarticletitle{Automatic code summarization via chatgpt: How far are we?}
\newblock \bibinfo{journal}{\emph{arXiv preprint arXiv:2305.12865}} (\bibinfo{year}{2023}).
\newblock


\bibitem[Sun et~al\mbox{.}(2024)]%
        {sun2024source}
\bibfield{author}{\bibinfo{person}{Weisong Sun}, \bibinfo{person}{Yun Miao}, \bibinfo{person}{Yuekang Li}, \bibinfo{person}{Hongyu Zhang}, \bibinfo{person}{Chunrong Fang}, \bibinfo{person}{Yi Liu}, \bibinfo{person}{Gelei Deng}, \bibinfo{person}{Yang Liu}, {and} \bibinfo{person}{Zhenyu Chen}.} \bibinfo{year}{2024}\natexlab{}.
\newblock \showarticletitle{Source code summarization in the era of large language models}.
\newblock \bibinfo{journal}{\emph{arXiv preprint arXiv:2407.07959}} (\bibinfo{year}{2024}).
\newblock


\bibitem[Tafreshipour et~al\mbox{.}(2024)]%
        {tafreshipour2024prompting}
\bibfield{author}{\bibinfo{person}{Mahan Tafreshipour}, \bibinfo{person}{Aaron Imani}, \bibinfo{person}{Eric Huang}, \bibinfo{person}{Eduardo Almeida}, \bibinfo{person}{Thomas Zimmermann}, {and} \bibinfo{person}{Iftekhar Ahmed}.} \bibinfo{year}{2024}\natexlab{}.
\newblock \showarticletitle{Prompting in the Wild: An Empirical Study of Prompt Evolution in Software Repositories}.
\newblock \bibinfo{journal}{\emph{arXiv preprint arXiv:2412.17298}} (\bibinfo{year}{2024}).
\newblock


\bibitem[Takebayashi et~al\mbox{.}(2023)]%
        {takebayashi2023exploratory}
\bibfield{author}{\bibinfo{person}{Taryn Takebayashi}, \bibinfo{person}{Anthony Peruma}, \bibinfo{person}{Mohamed~Wiem Mkaouer}, {and} \bibinfo{person}{Christian~D Newman}.} \bibinfo{year}{2023}\natexlab{}.
\newblock \showarticletitle{An exploratory study on the usage and readability of messages within assertion methods of test cases}. In \bibinfo{booktitle}{\emph{2023 IEEE/ACM 2nd International Workshop on Natural Language-Based Software Engineering (NLBSE)}}. IEEE, \bibinfo{pages}{32--39}.
\newblock


\bibitem[Tufano et~al\mbox{.}(2020)]%
        {tufano2020unit}
\bibfield{author}{\bibinfo{person}{Michele Tufano}, \bibinfo{person}{Dawn Drain}, \bibinfo{person}{Alexey Svyatkovskiy}, \bibinfo{person}{Shao~Kun Deng}, {and} \bibinfo{person}{Neel Sundaresan}.} \bibinfo{year}{2020}\natexlab{}.
\newblock \showarticletitle{Unit test case generation with transformers and focal context}.
\newblock \bibinfo{journal}{\emph{arXiv preprint arXiv:2009.05617}} (\bibinfo{year}{2020}).
\newblock


\bibitem[Wang et~al\mbox{.}(2024)]%
        {wang2024software}
\bibfield{author}{\bibinfo{person}{Junjie Wang}, \bibinfo{person}{Yuchao Huang}, \bibinfo{person}{Chunyang Chen}, \bibinfo{person}{Zhe Liu}, \bibinfo{person}{Song Wang}, {and} \bibinfo{person}{Qing Wang}.} \bibinfo{year}{2024}\natexlab{}.
\newblock \showarticletitle{Software testing with large language models: Survey, landscape, and vision}.
\newblock \bibinfo{journal}{\emph{IEEE Transactions on Software Engineering}} (\bibinfo{year}{2024}).
\newblock


\bibitem[Wang et~al\mbox{.}(2020)]%
        {wang2020generalizing}
\bibfield{author}{\bibinfo{person}{Yaqing Wang}, \bibinfo{person}{Quanming Yao}, \bibinfo{person}{James~T Kwok}, {and} \bibinfo{person}{Lionel~M Ni}.} \bibinfo{year}{2020}\natexlab{}.
\newblock \showarticletitle{Generalizing from a few examples: A survey on few-shot learning}.
\newblock \bibinfo{journal}{\emph{ACM computing surveys (csur)}} \bibinfo{volume}{53}, \bibinfo{number}{3} (\bibinfo{year}{2020}), \bibinfo{pages}{1--34}.
\newblock


\bibitem[Watson et~al\mbox{.}(2020)]%
        {watson2020learning}
\bibfield{author}{\bibinfo{person}{Cody Watson}, \bibinfo{person}{Michele Tufano}, \bibinfo{person}{Kevin Moran}, \bibinfo{person}{Gabriele Bavota}, {and} \bibinfo{person}{Denys Poshyvanyk}.} \bibinfo{year}{2020}\natexlab{}.
\newblock \showarticletitle{On learning meaningful assert statements for unit test cases}. In \bibinfo{booktitle}{\emph{Proceedings of the ACM/IEEE 42nd International Conference on Software Engineering}}. \bibinfo{pages}{1398--1409}.
\newblock


\bibitem[Wei et~al\mbox{.}(2022)]%
        {wei2022chain}
\bibfield{author}{\bibinfo{person}{Jason Wei}, \bibinfo{person}{Xuezhi Wang}, \bibinfo{person}{Dale Schuurmans}, \bibinfo{person}{Maarten Bosma}, \bibinfo{person}{Fei Xia}, \bibinfo{person}{Ed Chi}, \bibinfo{person}{Quoc~V Le}, \bibinfo{person}{Denny Zhou}, {et~al\mbox{.}}} \bibinfo{year}{2022}\natexlab{}.
\newblock \showarticletitle{Chain-of-thought prompting elicits reasoning in large language models}.
\newblock \bibinfo{journal}{\emph{Advances in neural information processing systems}}  \bibinfo{volume}{35} (\bibinfo{year}{2022}), \bibinfo{pages}{24824--24837}.
\newblock


\bibitem[Zamprogno et~al\mbox{.}(2022)]%
        {zamprogno2022dynamic}
\bibfield{author}{\bibinfo{person}{Lucas Zamprogno}, \bibinfo{person}{Braxton Hall}, \bibinfo{person}{Reid Holmes}, {and} \bibinfo{person}{Joanne~M Atlee}.} \bibinfo{year}{2022}\natexlab{}.
\newblock \showarticletitle{Dynamic human-in-the-loop assertion generation}.
\newblock \bibinfo{journal}{\emph{IEEE Transactions on Software Engineering}} \bibinfo{volume}{49}, \bibinfo{number}{4} (\bibinfo{year}{2022}), \bibinfo{pages}{2337--2351}.
\newblock


\bibitem[Zhang et~al\mbox{.}(2020)]%
        {zhang2020retrieval}
\bibfield{author}{\bibinfo{person}{Jian Zhang}, \bibinfo{person}{Xu Wang}, \bibinfo{person}{Hongyu Zhang}, \bibinfo{person}{Hailong Sun}, {and} \bibinfo{person}{Xudong Liu}.} \bibinfo{year}{2020}\natexlab{}.
\newblock \showarticletitle{Retrieval-based neural source code summarization}. In \bibinfo{booktitle}{\emph{Proceedings of the ACM/IEEE 42nd International Conference on Software Engineering}}. \bibinfo{pages}{1385--1397}.
\newblock


\bibitem[Zhang et~al\mbox{.}(2025)]%
        {zhang2025exploring}
\bibfield{author}{\bibinfo{person}{Quanjun Zhang}, \bibinfo{person}{Weifeng Sun}, \bibinfo{person}{Chunrong Fang}, \bibinfo{person}{Bowen Yu}, \bibinfo{person}{Hongyan Li}, \bibinfo{person}{Meng Yan}, \bibinfo{person}{Jianyi Zhou}, {and} \bibinfo{person}{Zhenyu Chen}.} \bibinfo{year}{2025}\natexlab{}.
\newblock \showarticletitle{Exploring automated assertion generation via large language models}.
\newblock \bibinfo{journal}{\emph{ACM Transactions on Software Engineering and Methodology}} \bibinfo{volume}{34}, \bibinfo{number}{3} (\bibinfo{year}{2025}), \bibinfo{pages}{1--25}.
\newblock


\bibitem[Zhang et~al\mbox{.}(2019)]%
        {zhang2019bertscore}
\bibfield{author}{\bibinfo{person}{Tianyi Zhang}, \bibinfo{person}{Varsha Kishore}, \bibinfo{person}{Felix Wu}, \bibinfo{person}{Kilian~Q Weinberger}, {and} \bibinfo{person}{Yoav Artzi}.} \bibinfo{year}{2019}\natexlab{}.
\newblock \showarticletitle{Bertscore: Evaluating text generation with bert}.
\newblock \bibinfo{journal}{\emph{arXiv preprint arXiv:1904.09675}} (\bibinfo{year}{2019}).
\newblock


\end{thebibliography}
